\documentclass[a4paper, amsfonts, amssymb, amsmath, reprint, showkeys, footinbib, twoside,superscriptaddress,floatfix,longbibliography]{revtex4-1}

\usepackage{graphicx} %
\graphicspath{{./}}
\usepackage{amsmath}
\usepackage{comment}

\usepackage{booktabs}
\usepackage{hyperref}
\usepackage{url}
\usepackage{adjustbox}
\usepackage{array}
\usepackage[english]{babel}

\usepackage{multirow}
\usepackage{hhline}
\usepackage{makecell}
\usepackage[table]{xcolor}

\newcommand{\figLabelCapt}[1]{\textbf{\MakeLowercase{{#1}}}}
\newcommand{\refSub}[2]{\hyperref[#2]{\ref{#2}\figLabelCapt{#1}}}
\newcommand{\figref}[1]{Fig.~\ref{#1}}
\newcommand{\figrefsub}[2]{Fig.~\refSub{#1}{#2}}

\newcommand{\br}[1]{\mathbf{r}}
\newcommand{\bk}[1]{\mathbf{k}}

\begin{document}

\title{Equivariant learning of a transferable three-dimensional classical density functional}

\author{Bingqing Cheng}
\email{bingqingcheng@berkeley.edu}
\affiliation{Department of Chemistry, UC Berkeley, California 94720, United States}
\affiliation{Chemical Sciences Division, Lawrence Berkeley National Laboratory, Berkeley, California, 94720, United States}
\affiliation{Bakar Institute of Digital Materials for the Planet, UC Berkeley, California 94720, United States}

\date{\today}

\begin{abstract}
\textbf{Abstract:}
Liquids exhibit collective behavior that depends sensitively on thermodynamic conditions, interfaces and confinement, yet predicting each new state commonly requires a separate atomistic simulation. Classical density functional theory offers a reusable variational description, but its central excess free-energy functional is generally unknown, and learned approximations for fluids have largely been developed in reduced-dimensional geometries. Here we show that this functional can be learned directly from fully three-dimensional canonical equilibrium density fields while preserving spatial symmetry and variational consistency, without free-energy or chemical-potential training labels. {A functional learned from coarse three-dimensional density fields remains predictive at finer resolution, separating the sampling demands of training from the spatial resolution of subsequent predictions.} A single learned functional transfers across temperatures, system sizes and statistical ensembles, and recovers the equation of state, liquid--vapor coexistence and interfacial broadening, none of which are used as training targets. Applied to complex three-dimensional geometries, it predicts the potential of mean force between colloids due to solvent mediation, and adsorption isotherms in an interconnected gyroid pore. These results demonstrate that molecular dynamics equilibrium density data can be converted into a transferable thermodynamic generator connecting microscopic liquid structure to response, phase behavior and collective phenomena.
\end{abstract}

\maketitle

\section{Introduction}

Liquids are paradigmatic many-body systems: even when interparticle
interactions are simple, collective structure and thermodynamics emerge from
the correlated configurations of all particles. These correlations connect
molecular organization to collective behavior across length scales and are
strongly reshaped by interfaces and confinement. Atomistic simulations can
resolve a chosen state with microscopic accuracy, but must be repeated across
thermodynamic conditions and geometries, motivating a reusable field-level
theory.

Density functional theory provides such a description by replacing the full
many-particle distribution with a variational theory of a one-body density.
Electronic density functional theory reformulates the quantum-mechanical
ground-state problem in terms of the electron density
~\cite{Hohenberg1964,Kohn1965}; classical density functional theory (cDFT) is
its statistical-mechanical analogue for the equilibrium number density
$\rho(\mathbf r)$~\cite{Evans1979}. Once known, the intrinsic free-energy
functional determines equilibrium density fields, thermodynamics and response
within a common framework.

The difficulty is that this functional is not known. In electronic DFT,
exchange and correlation must be approximated; in classical DFT, the ideal-gas
contribution is exact, but the interaction-induced excess free-energy
functional $F_{\mathrm{exc}}[\rho]$ remains unknown for realistic
three-dimensional fluids. Fundamental measure theory provides highly accurate
functionals for hard-sphere systems~\cite{Rosenfeld1989,Roth2010FMT}, whereas
attractive and long-ranged interactions are commonly treated using
perturbative or mean-field approximations~\cite{Evans2016Developments} whose
accuracy is system- and state-dependent. Constructing an accurate
$F_{\mathrm{exc}}[\rho]$ while preserving the variational structure of cDFT is
therefore the central theoretical challenge.

Machine learning offers a data-driven route to approximate this missing
functional using atomistic simulation data~\cite{SimonOettel2026Review}.
Earlier studies learned explicit excess free-energy functionals from
equilibrium planar density profiles or bulk pair
correlations~{\cite{LinOettel2019,Cats2021MLFreeEnergy,Dijkman2025PairMatching,Ram2025DDFT}}.
These approaches established both scalar-functional learning and local
free-energy representations as routes to transferable neural density
functionals~\cite{SammullerSchmidt2024LocalPair}. Related approaches
have constructed transferable functional approximations for molecular and
ionic liquids~\cite{Kelley2024Universal,Bui2025Ionic,BuiCox2026AbInitio}.
Neural operators instead learn forward and inverse potential--density maps
without constructing an explicit thermodynamic
functional~\cite{Pan2025NeuralOperators,Yang2025HighDimensional}.
Notably, neural functional theory directly learns the one-body direct
correlation functional
$c^{(1)}(\mathbf r;[\rho])=-\beta\,\delta F_{\mathrm{exc}}/\delta\rho(\mathbf r)$~\cite{Sammuller2023Neural}.
In its original formulation, grand-canonical Monte
Carlo (GCMC) samples of a three-dimensional hard-sphere fluid were generated under
randomized one-dimensional external fields $V_{\mathrm{ext}}(z)$. The imposed
chemical potential and external field provided reference $c^{(1)}(z)$ labels
through the equilibrium Euler--Lagrange balance, and a shared neural network
mapped local windows of the planar profile $\rho(z)$ to $c^{(1)}(z)$.
Subsequent work incorporated temperature and system size to construct thermal
functionals~\cite{Sammuller2025Coexistence}, treated the pair interaction as a
functional input~\cite{Kampa2025Metadensity}, generalized the representation to
binary mixtures~\cite{Robitschko2025Mixtures}, or inferred the chemical
potentials as latent variables so that canonical molecular dynamics (MD) data
could be used~\cite{Sammuller2026ChemicalPotential}. Neural cDFT has also been
extended from planar fields to two-dimensional inhomogeneities using
convolutional layers~\cite{Glitsch2025HigherDimensions},
and spherical symmetry~\cite{kampa2026spherical}.

Extending learned functionals to continuum fluids with unrestricted
three-dimensional inhomogeneity remains challenging. Such fields are essential
for solvation, nonplanar interfaces and interconnected confinement, where a
single wall-normal density profile is insufficient. Recent work demonstrated
cubic-equivariant direct-correlation learning for three-dimensional lattice
fluids~\cite{Weimar2026Cubic}. 
{A practical bottleneck is obtaining statistically converged
three-dimensional density fields from atomistic simulations, particularly at
fine spatial resolution.}
Here we address continuum fluids through a scalar
free-energy representation trained on fully three-dimensional equilibrium
density fields.

Here we introduce Equi-cDFT, a framework that combines an extensive
scalar free-energy representation, cubic-symmetry-adapted density descriptors
and canonical equilibrium training. Finite-range density
environments are encoded by cubic-symmetry-adapted representations, making the
scalar functional invariant and its functional derivatives equivariant under
rotations and reflections of the cubic lattice. 
{The density representation allows the same learned functional to be
evaluated at different grid spacings; we demonstrate accurate transfer
without retraining.}
To learn from canonical MD
density fields without chemical-potential labels, we train through local chemical-potential balance
and eliminate the unknown constant chemical
potential analytically. Minimization and functional differentiation of the
resulting scalar then provide equilibrium densities and direct correlation
functions within one variational model.

We demonstrate the approach for the Lennard--Jones fluid truncated and shifted
at $2.5\sigma$ (LJTS), a prototypical continuum model in which repulsive packing and
cohesive attraction generate rich phase and interfacial
behavior~\cite{Sammuller2025Coexistence}. Accurate reference equations of state
and coexistence data are available for this
model~\cite{Thol2015LJTSEOS,Vrabec2006LJTSCoexistence}, making it a controlled
but stringent test of whether one learned functional can recover consistent
structure and thermodynamics across diverse conditions.

We first test transfer across temperature, system size, statistical
ensemble, and {grid resolution} (see Methods section). We then ask whether physical information absent from training emerges
from the learned functional, including the equation of state,
liquid--vapor coexistence and interfaces. Finally, we apply the same
functional to two fully three-dimensional problems:
solvent-mediated interactions between colloids and adsorption in a periodic
gyroid pore under contrasting subcritical and supercritical conditions.

\section{Theory}

\subsection{Equivariant local representation of the excess free-energy functional}

\begin{figure*}
\centering
\includegraphics[width=\textwidth]{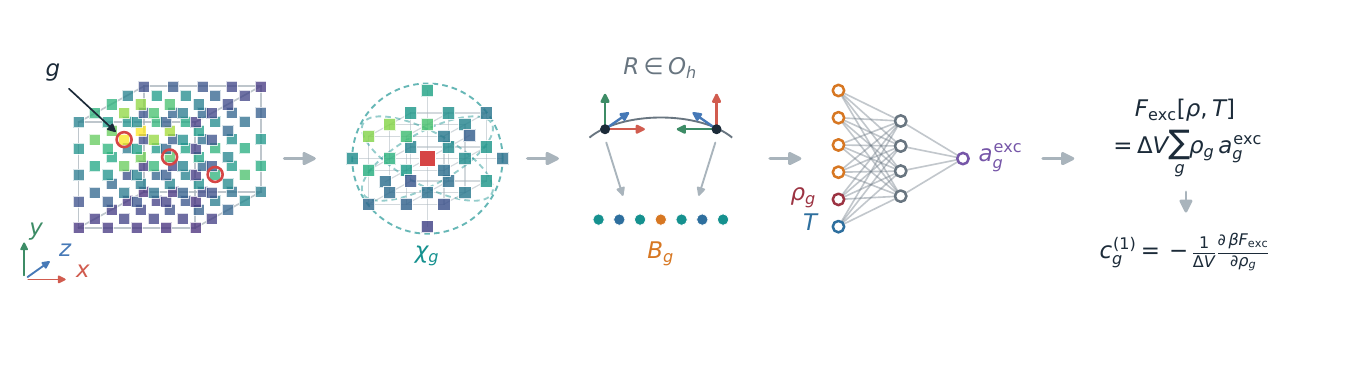}
\caption{\textbf{Equi-cDFT architecture.}
{The descriptor step converts the local density environment into
cubic-symmetry-adapted features, which are processed by a shared neural readout.}
Summing the local contributions produces an extensive excess free-energy
functional, whose derivative gives the one-body direct correlation field.
Here $g$ labels a density-grid point, $\chi_g$ is its finite spherical
environment, $R\in O_h$ denotes a rotation or reflection of the cubic
lattice, $\mathbf B_g$ is the resulting symmetry-adapted feature vector and
$T$ is the temperature. The readout produces the local excess free energy per
particle $a_g^{\mathrm{exc}}$; summing over the grid densities $\rho_g$ with
voxel volume $\Delta V$ gives $F_{\mathrm{exc}}[\rho,T]$.
$c_g^{(1)}$ is the one-body direct correlation, and
$\beta=(k_{\mathrm B}T)^{-1}$.}
\label{fig:algorithm-schematic}
\end{figure*}

At fixed temperature $T$, external potential $V_{\mathrm{ext}}$ and reservoir
chemical potential $\mu$, cDFT determines the equilibrium one-body density
$\rho(\mathbf r)$ by minimizing the grand-potential
functional~\cite{Evans1979}
\begin{equation}
\Omega[\rho]
=F_{\mathrm{id}}[\rho,T]+F_{\mathrm{exc}}[\rho,T]
+\int\!\mathrm d\mathbf r\,\rho(\mathbf r)
\bigl[V_{\mathrm{ext}}(\mathbf r)-\mu\bigr].
\label{eq:grand_potential}
\end{equation}
Here $F_{\mathrm{id}}=k_{\mathrm B}T\int\!\mathrm d\mathbf r\,
\rho\{\ln(\Lambda^3\rho)-1\}$ is the analytic ideal-gas contribution,
$V_{\mathrm{ext}}$ is the known external field, and $F_{\mathrm{exc}}$ is the
unknown excess free-energy functional containing the effects of interparticle
interactions.
For fixed-particle-number predictions, we minimize the learned
intrinsic functional plus the external-field contribution subject to the
constraint $\int\rho\,\mathrm d\mathbf r=N$, with a spatially constant
Lagrange multiplier enforcing this constraint. Canonical MD densities can be used
to approximate the intrinsic functional conventionally defined in
grand-canonical cDFT, invoking ensemble equivalence in the thermodynamic limit.

Equi-cDFT learns $F_{\mathrm{exc}}[\rho,T]$ as a function of temperature and a
three-dimensional density field represented on a regular grid with voxel
volume $\Delta V = h^3$, {where $h$ is the grid spacing}. Short-ranged interactions motivate representing each local
contribution using a finite neighborhood, but do not imply that fluid
correlations are strictly local. We therefore treat the finite receptive field
as a modeling approximation that enables transfer across system sizes and test
its adequacy in the benchmarks below. As illustrated in
\figref{fig:algorithm-schematic}, the local environment $\chi_g$ contains the
density within a finite neighborhood of grid point $g$. We approximate the
excess functional as
\begin{equation}
F_{\mathrm{exc}}[\rho,T]
=\Delta V\sum_g\rho_g\,
a_{\mathrm{exc}}(\chi_g,T),
\label{eq:local_energy_decomposition}
\end{equation}
using the same local free-energy map $a_{\mathrm{exc}}$ at every grid point.
This shared map makes the functional extensive and allows it to transfer
between system sizes.
{In the implementation, this contribution combines two neural readouts:
a local-density-approximation baseline depending on density of the voxel and
temperature, and an environment-dependent correction using the CACE features.
Their outputs are added before assembling and differentiating the scalar
functional.}

Within this decomposition, we use a symmetry-adapted representation of the
local three-dimensional environments $\chi_g$. A rotation or reflection belonging to
the symmetry group of the cubic grid must leave the scalar local free energy
$a_{\mathrm{exc}}(\chi_g,T)$ unchanged. We construct this representation by
adapting the Cartesian atomic cluster expansion
(CACE)~\cite{cheng2024cartesian,drautz2019atomic} to a density grid. For a
target grid point $\mathbf r_g$, let $\mathbf q=(q_x,q_y,q_z)$ denote the
integer offsets of neighboring voxels within the cutoff. The Cartesian moments
are
\begin{equation}
A_{\boldsymbol\ell}(\mathbf r_g)
=\sum_{{0<|\mathbf q|\leq q_{\mathrm{cut}}}}
\rho_{g+\mathbf q}\,{w_\alpha(\mathbf q)}\,
q_x^{\ell_x}q_y^{\ell_y}q_z^{\ell_z},
\label{eq:lattice_cace_moments}
\end{equation}
where $\boldsymbol\ell=(\ell_x,\ell_y,\ell_z)$ specifies the Cartesian powers,
{and $w_\alpha(\mathbf q)$ is a normalized Gaussian radial basis as specified
in Methods.}
These moments are covariant under the cubic point group $O_h$, which consists
of the $3!$ axis permutations combined with $2^3$ independent axis
reflections, giving 48 operations in total. Under a group operation
$\mathcal R\in O_h$, a moment transforms as
\begin{equation}
A_{\boldsymbol\ell}
\longrightarrow
s_{\boldsymbol\ell}(\mathcal R)
A_{\mathcal R\boldsymbol\ell},
\label{eq:lattice_cace_moment_transform}
\end{equation}
where $\mathcal R\boldsymbol\ell$ denotes the permutation of the power tuple
induced by $\mathcal R$, and $s_{\boldsymbol\ell}(\mathcal R)=\pm1$ is the
parity acquired from reflections of axes carrying odd powers.

For $K=(\boldsymbol\ell_1,\ldots,\boldsymbol\ell_\nu)$, the corresponding
invariant feature is obtained by summing the transformed products over
$O_h$,
\begin{equation}
B_K^{(\nu)}(\mathbf r_g)
=
\sum_{\mathcal R\in O_h}
\prod_{m=1}^{\nu}
s_{\boldsymbol\ell_m}(\mathcal R)
A_{\mathcal R\boldsymbol\ell_m}(\mathbf r_g).
\label{eq:lattice_cace_invariants}
\end{equation}
The invariant features $B_K^{(\nu)}$, together with the center density and
temperature, are passed to a shared scalar readout to obtain
$a_{\mathrm{exc}}(\chi_g,T)$ at every grid point.

{Importantly, the use of the CACE representation effectively projects the density field on the grid onto a set of basis functions. 
This representation allows the same learned functional to be evaluated on different grids, with physical length
scales held fixed and the discretization adjusted consistently.
This is analogous to the grid resolution transferability of
neural-operator learning~\cite{Kovachki2023NeuralOperator,Li2021FourierOperator}.
In Methods, we show that the Equi-cDFT model trained using native grid spacing $h_0=0.5\sigma$ is transferable to finer grids $0.25\sigma$ and $0.125\sigma$,
as well as coarser grid $1\sigma$.}

\subsection{Derivative learning from local chemical-potential balance}

We learn the excess free energy through its local density gradient, computed by
automatic differentiation,
\begin{equation}
c^{(1)}_g([\rho],T)
=-\frac{\beta}{\Delta V}
\frac{\partial F_{\mathrm{exc}}[\rho,T]}
{\partial\rho_g}.
\label{eq:c1}
\end{equation}
Here $c^{(1)}$ is the one-body direct correlation functional.
Neural cDFT has been developed through direct learning of
$c^{(1)}$~\cite{Sammuller2023Neural,Sammuller2025Coexistence}.
{Learning a scalar excess functional through its first derivative was
previously demonstrated for planar density profiles by Samm{\"u}ller and
Schmidt~\cite{SammullerSchmidt2024LocalPair}.
With the same inhomogeneous training data, their scalar-functional and
direct-correlation representations produced comparably accurate equilibrium
density profiles.}

For an equilibrium density field in either the canonical or grand-canonical
ensemble, define the dimensionless local chemical potential
\begin{equation}
\beta \mu^{\mathrm{loc}}_g
=\ln\!\left(\Lambda^3\rho_g\right)
+\beta V_{\mathrm{ext},g}
-c^{(1)}_g([\rho],T).
\label{eq:local_chemical_potential}
\end{equation}
The equilibrium condition requires this quantity to be independent of
position. A spatial imbalance would drive a redistribution of density;
equilibrium is reached when that driving force vanishes
everywhere~\cite{Evans1979}.

Canonical equilibrium training requires treating the unknown chemical
potential of each reference field. Samm{\"u}ller and Schmidt addressed this by
jointly learning the direct-correlation functional and a chemical potential
for each simulation~\cite{Sammuller2026ChemicalPotential}. We use the same
equilibrium balance, but eliminate the per-field constant analytically: for
the spatial weighting used here, its least-squares optimum is the mean
predicted local chemical potential. Substituting this optimum gives the
centered loss
\begin{equation}
\mathcal L^{\Delta\mu}_{{s}}
=\sum_g\left|
\beta \mu^{\mathrm{loc}}_g
-\overline{\beta \mu^{\mathrm{loc}}}
\right|^2,
\label{eq:chemical_potential_balance_loss}
\end{equation}
where both the sum and the average are taken over the reliably sampled region
of the density field {$s$. During training,
the losses are accumulated over all fields in a batch.}
If the reference data include a chemical-potential label,
the spatial mean in Eq.~\eqref{eq:chemical_potential_balance_loss} can instead
be replaced
by {$\beta\mu_{\mathrm{ref}}$}, anchoring the absolute chemical-potential
reference.

{Canonical balance leaves a temperature-dependent gauge freedom}: the loss is unchanged under
\begin{equation}
F_{\mathrm{exc}}[\rho,T]
\longrightarrow F_{\mathrm{exc}}[\rho,T]+b(T)N.
\label{eq:canonical_free_energy_gauge}
\end{equation}
{At each temperature, the linear gauge shifts all chemical potentials by
the same constant $b(T)$, leaving chemical-potential differences between isothermal states
unchanged. Fixed-particle-number density predictions, pressure and $c^{(2)}$
are also unaffected. 
If an absolute chemical-potential reference is required, it can be fixed at
each temperature by imposing the exact dilute-limit condition
$c^{(1)}[\rho=0,T]=0$, which the present readout does not enforce automatically,
or by using a reference chemical potential.}

During training, independent equilibrium density fields collected from
canonical molecular dynamics under randomly selected external fields are
supplied. For inference, $V_{\mathrm{ext}}$ is given and the total free
energy is minimized with respect to the grid densities, either at fixed total
particle number or fixed chemical potential.

\section{Results}

\subsection{Equation of state, phase behavior, and interfaces}

{Throughout the following tests, we evaluate the functional trained at
$h_0=0.5\sigma$ on both $h=0.5\sigma$ and $h=0.25\sigma$ grids without
retraining. The grid-transfer procedure is described in Methods.}

\begin{figure*}
\centering
\includegraphics[width=0.85\textwidth]{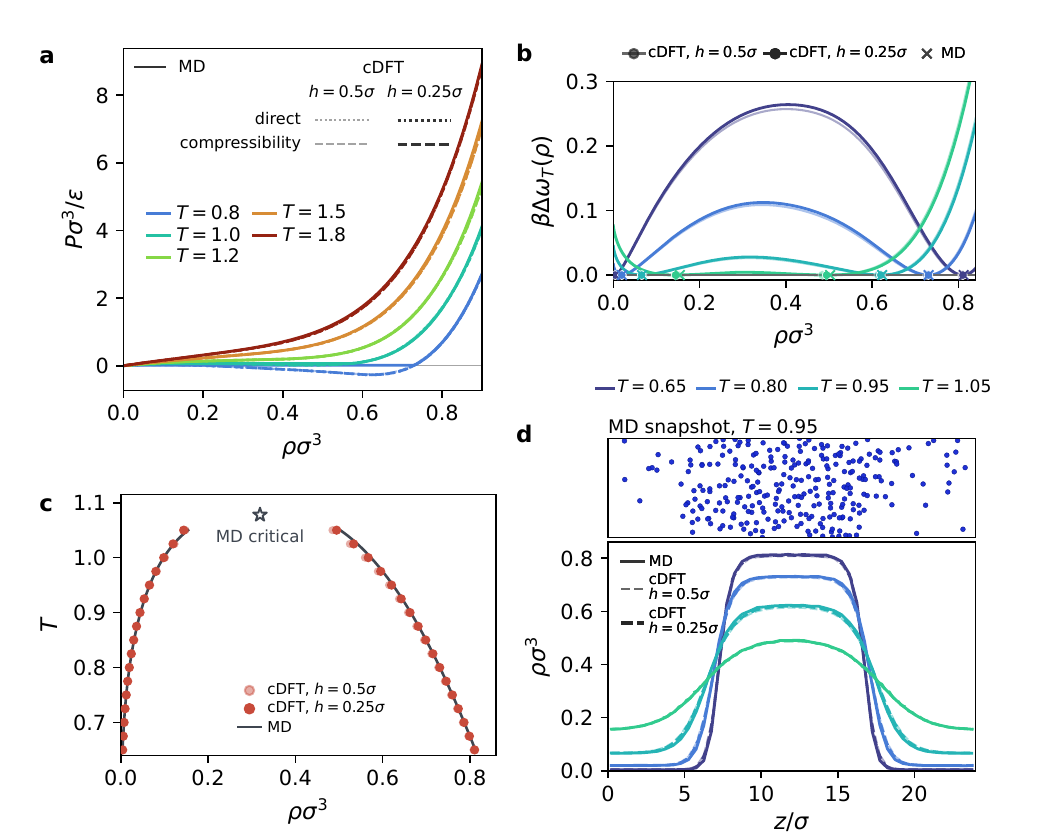}
\caption{\textbf{Equation of state, phase behavior, and interfaces from the
learned free-energy functional.}
{For cDFT predictions, lighter colors show results based on grid spacing $h=0.5\sigma$ and the darker colors are from $h=0.25\sigma$.}
\textbf{a} Pressure--density isotherms.
Dotted curves are evaluated directly from the learned scalar free energy,
dashed curves are obtained by compressibility-route integration, and solid
curves are evaluations of the MD-derived reference EOS~\cite{Thol2015LJTSEOS}.
\textbf{b} Homogeneous free-energy density relative to the predicted common
tangent, $\beta\Delta\omega_T(\rho)
=\beta[f_T(\rho)-\mu_{\mathrm{coex}}\rho+P_{\mathrm{coex}}]$.
The horizontal zero line represents the tangent; colored circles mark the
predicted vapor and liquid contact densities, and crosses mark the reference
EOS coexistence densities~\cite{Thol2015LJTSEOS}. 
\textbf{c} Liquid--vapor coexistence. Unconnected red circles are common-tangent
densities from the learned homogeneous free energy through $T=1.05$.
The gray branches show the reference EOS; the open
star is the direct-simulation critical-point estimate~\cite{Thol2015LJTSEOS,Vrabec2006LJTSCoexistence}.
\textbf{d} An atomistic MD slab at $T=0.95$ (top), and planar density profiles
at $T=0.65$, 0.80, 0.95 and 1.05 (temperature colors are
shared with \textbf{b}). Solid curves are averaged MD
profiles; dashed curves are fixed-$N$ Equi-cDFT solutions in the same geometry.}
\label{fig:bulk-response-phase}
\end{figure*}

This benchmark tests whether the learned model behaves as a
thermodynamic functional rather than merely reproducing equilibrium fields
under external potentials. Neither free energies, bulk response, pressure,
phase coexistence nor free interfaces entered the training loss. We therefore
ask whether the same scalar functional recovers bulk phase separation, its
equation of state and the structure of a liquid--vapor interface.

We first evaluate the Helmholtz free-energy density
$f_T(\rho)=(F_{\mathrm{id}}+F_{\mathrm{exc}})/V$ for {homogeneous} fluids. 
This scalar free energy gives the chemical potential
$\mu_T(\rho)=\partial f_T/\partial\rho$ and the pressure
\begin{equation}
P(\rho,T)
=\rho\mu_T(\rho)-f_T(\rho),
\label{eq:results_direct_pressure}
\end{equation}
with the canonical $b(T)N$ gauge cancelled.
The cDFT pressures from this direct route at different density and {temperature} conditions are shown as dotted curves in \figrefsub{a}{fig:bulk-response-phase}.

An alternative, compressibility pressure route uses the
two-body direct correlation function obtained by differentiation:
\begin{equation}
c^{(2)}(\mathbf r_g,\mathbf r_{g'};[\rho],T)
=\frac{1}{\Delta V}
\frac{\partial c^{(1)}(\mathbf r_g;[\rho],T)}
{\partial \rho(\mathbf r_{g'})}.
\label{eq:learned_second_direct_correlation}
\end{equation}
Reciprocity under $g\leftrightarrow g'$ follows from this common
scalar origin. Unlike approaches using pair-correlation
matching~\cite{SammullerSchmidt2024LocalPair,Dijkman2025PairMatching,Ram2025DDFT},
we do not supervise this second derivative. In a homogeneous fluid,
$c^{(2)}$ depends only on separation. Its zero-wavevector transform
$\widehat c^{(2)}(0)=\Delta V\sum_g c^{(2)}(\mathbf r_g)$ determines the
inverse isothermal compressibility through
\begin{equation}
\frac{\partial\,\beta P}{\partial\rho}
=1-\rho\widehat c^{(2)}(0;\rho,T).
\label{eq:results_compressibility_route}
\end{equation}
Integrating from the ideal-gas limit gives
\begin{equation}
\beta P(\rho,T)
=\int_0^\rho\!\mathrm d\rho'\,
\left[1-\rho'\widehat c^{(2)}(0;\rho',T)\right].
\label{eq:results_compressibility_pressure}
\end{equation}
In \figrefsub{a}{fig:bulk-response-phase}, the dotted direct-scalar
and dashed compressibility-route curves nearly coincide, agreeing to the
numerical accuracy of the density integration. This checks consistency
between the scalar and its derivatives; the comparison with the MD-derived
reference EOS provides the separate test of physical
accuracy~\cite{Thol2015LJTSEOS}.

At the subcritical temperatures $T=0.8$ and 1.0, the learned homogeneous
isotherms develop a van der Waals loop, including a mechanically unstable
region with $\partial P/\partial\rho<0$; at the lowest temperature the
predicted pressure also becomes negative. This loop continues the homogeneous
branch through the two-phase density range, whereas the stable reference EOS
replaces that region by the constant coexistence-pressure segment. Its
emergence without pair-response or pressure supervision shows that the learned
functional captures the nonconvex thermodynamics underlying liquid--vapor
separation~\cite{Sammuller2025Coexistence}.

As another manifestation,
at subcritical temperatures the free-energy density
$f_T(\rho)$ is nonconvex, which permits a common tangent touching the curve at a vapor
density and a liquid density. The tangent has slope $\mu_{\mathrm{coex}}$ and
intercept $-P_{\mathrm{coex}}$, expressing equality of chemical potential and
pressure between the two phases. To display this construction at several
temperatures, \figrefsub{b}{fig:bulk-response-phase} subtracts each predicted
tangent:
\begin{equation}
\beta\Delta\omega_T(\rho)
=\beta\left[f_T(\rho)-\mu_{\mathrm{coex}}\rho+P_{\mathrm{coex}}\right].
\label{eq:results_common_tangent_distance}
\end{equation}
The two contacts lie at zero, while the positive free-energy excess
between them shows why a homogeneous intermediate-density state is
unfavorable relative to bulk phase separation. The colored circles are
predicted tangent densities; the crosses mark the corresponding reference
MD EOS densities~\cite{Thol2015LJTSEOS}.

At a mean density between the two contacts, partitioning the fluid
into vapor and liquid lowers its bulk free energy relative to a uniform
state. The tangent gap quantifies this penalty per unit volume. With
increasing temperature, the contact densities approach one another and the
penalty decreases, reflecting the weakening distinction between liquid and
vapor. Note that this bulk free-energy-density gap is not a nucleation barrier: it
contains neither the geometry nor the interfacial cost of forming a droplet.

Repeating the common-tangent construction across temperature gives
the coexistence envelope in \figrefsub{c}{fig:bulk-response-phase}, which is in good agreement with the MD reference~\cite{Thol2015LJTSEOS,Vrabec2006LJTSCoexistence}.
Recovering bulk coexistence does not by itself guarantee an
accurate interface. An interface also probes how the functional balances
the tendency toward phase separation against the cost of spatially varying
density. Its temperature-dependent broadening therefore tests the spatial
coupling in the learned functional beyond the homogeneous equation of
state~\cite{Sammuller2025Coexistence}.
\figrefsub{d}{fig:bulk-response-phase} compares equilibrium
liquid--vapor profiles at the same cell geometry and particle number as the MD
reference. It captures the progressive interfacial broadening with increasing
temperature. The MD profiles are averaged after removing rigid slab
translation, but retain the effects of lateral interfacial fluctuations
permitted by the finite simulation cell. Equi-cDFT predicts a stationary
one-body density rather than explicitly sampling these fluctuating interfaces;
agreement therefore tests the averaged profile, not the capillary-wave
spectrum.

{The two grid resolutions give closely consistent pressures, coexistence
densities and interfacial profiles (\figref{fig:bulk-response-phase}). This
agreement supports transfer of both the homogeneous thermodynamics and the
interfacial structure beyond the training grid, and without training on these observables.}

\begin{figure*}[!t]
\centering
\includegraphics[width=\textwidth]{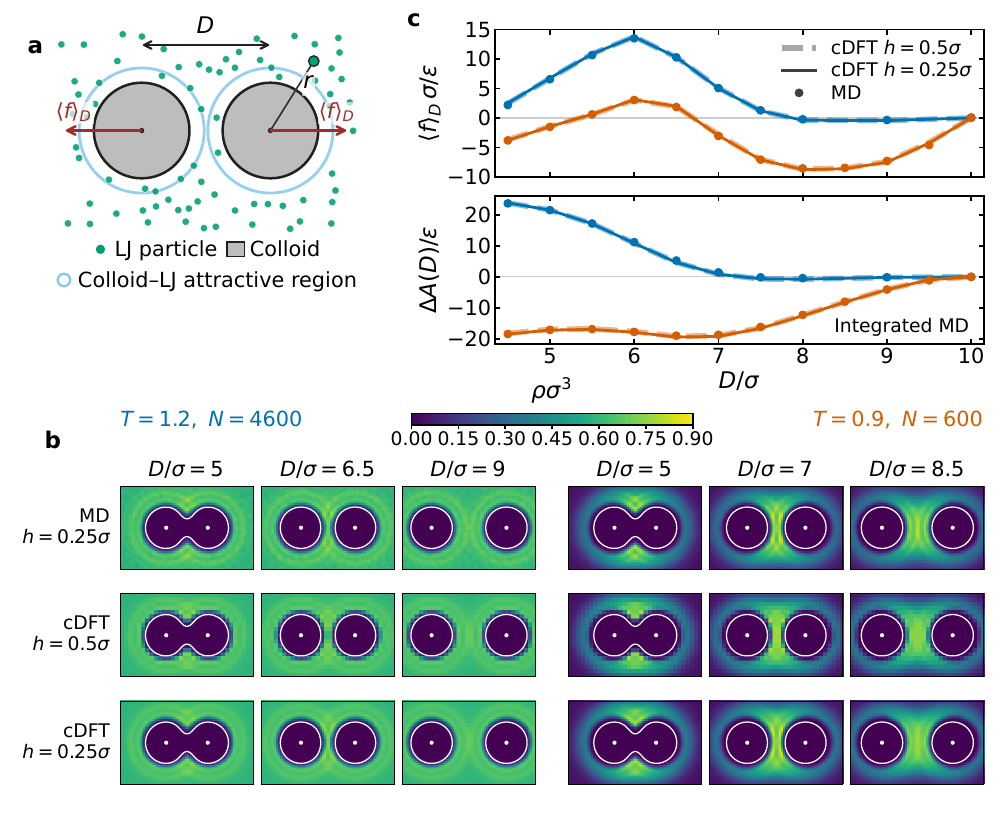}
\caption{\textbf{Solvent-mediated interactions between two fixed colloids.}
\textbf{a} Schematic of two colloids at center-to-center separation $D$
immersed in an LJ solvent. Pale blue rings mark the colloid--LJ attractive
regions, not a direct colloid--colloid interaction. Here $r$ is the distance
from a colloid center to a solvent particle. Outward red arrows define the
positive, repulsive generalized force $\langle f\rangle_D$; negative force is
attractive.
\textbf{b} Central $1\sigma$-thick slab averages of the number density from
MD and Equi-cDFT simulations {at grid spacings $h=0.5\sigma$ or $h=0.25\sigma$} using the same cubic $L=20\sigma$ box.
Both conditions use the same density scale, and white contours mark
$V_{\mathrm{ext}}/\epsilon=4$; white dots mark the colloid centers.
\textbf{c} Mean force (upper) and relative canonical free energy {$\Delta A(D)$}
(lower) for the two conditions, identified by blue and orange text.
Lines are Equi-cDFT predictions {at $h=0.5\sigma$ (lighter dashed) and $h=0.25\sigma$ (opaque solid)}; filled, unconnected circles are direct MD
forces or their trapezoidal integrals. For each condition separately, {$\Delta A(D)=A_N(D)-A_N(10\sigma)$}. Error bars for MD results are
sampling standard errors.
{Equi-cDFT force curves are evaluated from the predicted densities using
the Feynman--Hellmann expression, whereas the free-energy curves are evaluated
directly from the scalar functional. MD force points are direct ensemble
averages; the corresponding free-energy differences are obtained by
integrating those forces.}}
\label{fig:colloid-bridging}
\end{figure*}

\subsection{Solvent-mediated colloid bridging}

Nearby immersed objects can reorganize a solvent collectively and thereby
interact even without a direct mutual potential. Such solvation forces are a
longstanding application of traditional cDFT~\cite{EvansMarconi1987SolvationForces}. 
Here exclusion by two colloidal
cores {with steep repulsion}, which have no direct mutual forces,
competes with adsorption in their attractive shells. 
Changing their
separation causes the three-dimensional solvation environments to overlap,
form a bridge, and eventually recover two independent
solvation shells. Predicting the resulting interaction tests whether the
learned functional can translate collective solvent structure into a mean force. 
We compare a dilute subcritical solvent ($T=0.9$, $N=600$ LJ
particles) and a dense supercritical solvent ($T=1.2$, $N=4600$ LJ particles)
in the same periodic box of side $L=20\sigma$, using identical colloid fields.
The simulation parameters are given in Methods.
{Additional results for colloids with softer and harder
repulsive potentials are shown in Supplementary Fig.~S1.}

\figrefsub{b}{fig:colloid-bridging} compares {MD reference density fields
with Equi-cDFT predictions}.
For the dense supercritical solvent ($T=1.2$, $N=4600$), the
depleted regions around the two cores merge into a neck at
$D/\sigma=5.0$, which then disappears near $D/\sigma=6.5$. 
{In the dilute subcritical solvent, the depleted neck at small separation
gives way to an enriched intercolloid region at $D=7\sigma$, which remains
connected at $D=8.5\sigma$. The dense and dilute solvents therefore exhibit
contrasting reorganizations of the fluid between the colloids.}
{The finer grid better resolves the near-colloid density modulations
observed in MD (\figrefsub{b}{fig:colloid-bridging}).} 

With no direct potential, the colloid--colloid interaction
originates entirely from the solvent.
The solvent-mediated mean force between the colloids can be obtained
directly from the equilibrium solvent density through
the classical Feynman--Hellmann identity~\cite{Feynman1939Forces}:
\begin{equation}
\left\langle f \right\rangle_D
=-\Delta V\sum_g\rho_{D,g}
\frac{\partial V_{\mathrm{ext},g}(D)}{\partial D},
\label{eq:colloid_feynman_hellmann_grid}
\end{equation}
where $\rho_{D,g}$ is the equilibrium density at grid point $g$ and $\Delta V$
is the voxel volume.
Thus the learned functional predicts the interaction through its equilibrium
density alone; negative force denotes attraction and positive force repulsion.
Because the force is a weighted integral over the complete density field, it
is more demanding than pointwise density agreement: small but spatially
correlated density errors can produce appreciable force errors.
The upper plot in \figrefsub{c}{fig:colloid-bridging} compares
Equi-cDFT mean forces with direct MD ensemble averages. The learned functional
captures the overall force trends in both conditions, including the {repulsive maximum in the dense solvent and the stronger attraction
in the dilute solvent}.

The {potential of mean force} can be obtained from the separation-dependent canonical
free energy directly. We denote its difference from the $D=10\sigma$ state by
\begin{equation}
{\Delta A(D)}=A_N(D)-A_N(10\sigma),
\label{eq:colloid_relative_canonical_energy}
\end{equation}
where $A_N$ is the canonical functional at fixed $N$. 
Its slope gives the mean force,
{$\langle f\rangle_D=-\mathrm d\Delta A/\mathrm dD$}.
The force reveals the local tendency to approach or separate,
whereas the free-energy profile identifies favored and unfavorable separations. 
The unknown $b(T)N$ gauge in Eqn.~\eqref{eq:canonical_free_energy_gauge} cancels between states at the same $N$ and $T$, so
no gauge fit is needed. 
{For the two grid resolutions, both the mean forces
and the free-energy differences agree well,
demonstrating robustness of the predicted interaction to grid refinement.}

{The dense solvent produces a pronounced free-energy penalty for close
approach, whereas the dilute solvent favors a finite-separation basin near
$D=7\sigma$. The strongest attractive force occurs at larger separations and
drives approach toward this basin. The directly evaluated scalar free-energy differences also
agree with the integrated MD forces at both grid resolutions, testing the
learned functional beyond its local derivative.}
The contrasting density structures suggest competing roles for exclusion and adsorption: in the dense solvent, close approach joins the
depleted regions around the cores and subsequently reorganizes the opposing
solvation layers, whereas the dilute solvent exhibits pronounced adsorption
and enrichment between the colloids. This interpretation connects the
present example to established adsorption-mediated bridging
physics~\cite{Archer2005Bridging}.

\subsection{Adsorption in a porous gyroid}

\begin{figure*}[!t]
\centering
\includegraphics[width=\textwidth]{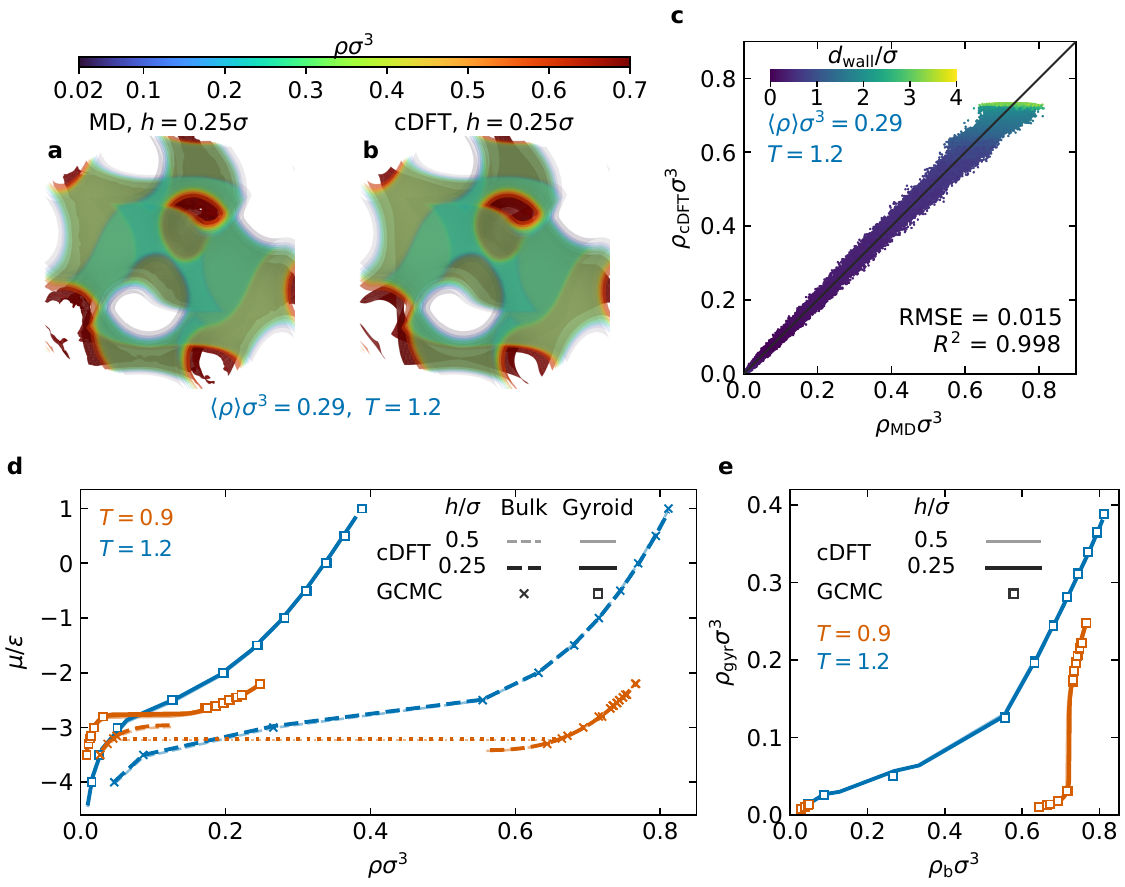}
\caption{\textbf{Gyroid-pore structure and adsorption.}
\textbf{a,b} Equilibrium MD and cDFT density fields at $T=1.2$ and the
matched isosurface levels, for whole-box mean density
$\langle\rho\rangle\sigma^3=0.29$. 
\textbf{c} {cDFT density versus the MD reference; both
densities use a grid spacing of $0.25\sigma$. Points are}
colored by the minimum periodic distance $d_{\mathrm{wall}}$ to the nominal
pore wall $V_{\mathrm{ext}}/\epsilon=4$.
\textbf{d} Chemical potential versus whole-box density at $T=1.2$ (blue)
and $T=0.9$ (orange). {Lighter/thinner and opaque/thicker lines denote
$h=0.5\sigma$ and $h=0.25\sigma$, respectively; dashed/solid curves denote
bulk/pore states. Dotted orange segments mark the model bulk common-tangent
construction.}
\textbf{e} Adsorption isotherms formed by matching bulk and pore states at
equal chemical potential.
Both bulk and pore densities use the full simulation-box volume.}
\label{fig:gyroid-adsorption}
\end{figure*}

Adsorption couples two distinct physical problems: the local organization of a
fluid under confinement and the thermodynamic partitioning of particles
between the pore and an external reservoir. A predictive density functional
must describe both with the same free-energy landscape. Classical DFT is
widely used for this purpose, including machine-learning studies of gas
solubility in nanopores~\cite{Qiao2020Nanopores} and fully three-dimensional
calculations of adsorbate structure in metal--organic and covalent--organic
frameworks~\cite{Soares2023MOF5,Stierle2024ThreeDimensional}. The gyroid makes
the test especially demanding: its continuously curved, interconnected pore
has no unique wall-normal coordinate and is topologically unlike the external
fields used for training. Its geometry and potential are specified in Methods.
{Additional results for pores with softer and harder
repulsive walls are shown in Supplementary Fig.~S2.}

The gyroid wall drives the density from nearly complete depletion near
the excluded region to dense, connected domains in the pore interior. At the
fixed loading shown in \figrefsub{a}{fig:gyroid-adsorption}--
\figrefsub{c}{fig:gyroid-adsorption}, Equi-cDFT reproduces this full
three-dimensional morphology, including the curved depletion layer and the
spatially varying accumulation throughout the interconnected pore. Coloring
the density comparison by distance from the wall shows that the agreement
extends from the strongly perturbed interfacial region into the pore interior.
This is therefore not merely recovery of the mean loading, but of how that
loading distributes across local environments with different wall distances
and curvatures.

Adsorption is governed by chemical equilibrium between the pore
and its reservoir. 
\figrefsub{d}{fig:gyroid-adsorption} compares the bulk and pore branches at
a supercritical ($T=1.2$) and a subcritical temperature ($T=0.9$): the horizontal axis is whole-box density and the vertical
axis is chemical potential. 
{The two grid resolutions yield closely consistent values.}
To align Equi-cDFT chemical potentials with the GCMC results, at each temperature {and grid spacing}, one 
chemical-potential offset calibrated from bulk references is shared with the pore.

On each branch, a horizontal line at a chosen chemical
potential in \figrefsub{d}{fig:gyroid-adsorption} selects a bulk density {on the chosen reservoir branch}
$\rho_{\mathrm b}$ and a pore density $\rho_{\mathrm{gyr}}$. Plotting the
latter against the former gives the adsorption isotherm in
\figrefsub{e}{fig:gyroid-adsorption}. 
Note that {in this matching procedure}, the chemical-potential offset cancels out and has no influence on the resulting adsorption isotherm.

At $T=0.9$, the pore remains weakly populated even along the dense
bulk-reservoir branch before filling sharply at higher chemical potential.
This can be rationalized by capillary evaporation from the purely repulsive wall~\cite{Roth2006}. 
At $T=1.2$, above the bulk critical point, filling is continuous rather than a
capillary phase transition. It is nevertheless strongly nonlinear: the pore
remains weakly populated in a dilute reservoir and fills rapidly as the
reservoir enters the dense-fluid regime. Capturing the distinct {subcritical} and {supercritical} adsorption behaviors together
with the confined density field shows that the learned functional transfers
not only across geometry, but thermodynamic conditions.

\section{Discussion}

Equi-cDFT extends learned scalar free-energy functionals to continuum
fluids with unrestricted three-dimensional density variations. Its
contribution lies in combining a local, extensive functional with
cubic-symmetry-adapted representations and canonical equilibrium training,
and demonstrating their transfer to spatially complex fluid environments.
The resulting model connects equilibrium structure, response and collective
behavior through one common scalar functional. 

The benchmarks demonstrate transfer across temperature, system size and
statistical ensemble (\figrefsub{a}{fig:combined-benchmarks} and
\figrefsub{b}{fig:combined-benchmarks}). Without retraining, it transfers to
larger cells (\figrefsub{c}{fig:combined-benchmarks}) and from canonical
training to grand-canonical chemical-potential differences
(\figrefsub{d}{fig:combined-benchmarks}). These tests probe more than
interpolation at reference densities: the inverse calculations explore the
constrained free-energy landscape, while the larger-cell test directly
examines the locality and extensivity assumptions. {The} consistent scalar and compressibility
pressure routes, direct
free-energy common tangents, phase coexistence and interfacial broadening
(\figref{fig:bulk-response-phase}) provide a
complementary test of emergent response and thermodynamics, because none of
these quantities entered the loss.

{A practical bottleneck for three-dimensional neural cDFT is obtaining
statistically converged MD density fields at fine spatial resolution: smaller
histogram bins require more sampling. By projecting local density
environments onto a common CACE representation at fixed physical scales,
Equi-cDFT separates the resolution needed for training from that used for
subsequent predictions. One can train on relatively coarse data and evaluate
the functional at finer resolution.
\figref{fig:grid-resolution-transfer} demonstrates inference at grid spacings
down to $0.125\sigma$, four times finer than the training grid.
Figures~\ref{fig:bulk-response-phase}--\ref{fig:gyroid-adsorption} further show
that refinement preserves the principal thermodynamic and application
predictions.
In the colloid example
(\figref{fig:colloid-bridging}), refinement better resolves near-surface
density modulations that are blurred on the coarse grid, while accurately
predicting the solvent-mediated mean force.}

{The learning-curve comparison  (\figref{fig:training-data-efficiency}) further supports the practical value of
the CACE representation: it reduces the amount of training data needed to
reach a given validation accuracy relative to the tested non-equivariant
voxel network. A scalar representation
enforces consistency between free energies and their derivatives, but does
not ensure their physical accuracy. The voxel network also has this scalar
structure, yet gives poorer bulk and interfacial predictions (Supplementary
Fig.~S3). This comparison highlights the importance of the local-environment
representation for learning a physically predictive functional from limited
data.}

The applications extend this generalization to full three-dimensional density
fields and to observables that were not supplied as labels. In the
two-colloid calculation, Equi-cDFT
predicts the non-monotonic solvent-mediated
force through the Feynman--Hellmann relation
(\figref{fig:colloid-bridging}),
and also provides canonical
free-energy differences directly. 
It resolves contrasting
solvent-depleted and solvent-enriched three-dimensional structures and their
associated free-energy profiles under the two thermodynamic conditions.
The gyroid is an even stronger geometric test:
its interconnected curved pore cannot be represented by a single wall-normal
coordinate. The same functional connects three-dimensional confined
structure to reservoir-controlled loading and predicts adsorption isotherms
(\figref{fig:gyroid-adsorption}), spanning continuous supercritical filling
and a sharper subcritical response.
{These predictions remain accurate for substantially steeper repulsive
confinement (Supplementary Figs.~S1 and S2).}
These are emergent
predictions of one learned thermodynamic generator: generalized forces arise
by differentiating its minimum, while adsorption arises by matching
independently computed bulk and confined states at equal chemical potential.

The present LJTS functional is interaction-specific, but the construction
suggests several direct extensions. One is to dynamical cDFT simulations, in
which the learned equilibrium functional supplies the adiabatic thermodynamic
driving force for time-dependent density evolution. A recent study showed that
a learned free-energy functional can be used without retraining in overdamped dynamical
density functional theory and grand-canonical gradient flows, with accuracy
set by the underlying adiabatic approximation~\cite{Ram2025DDFT}.
{Genuine nonequilibrium, superadiabatic
forces require additional functional information, as addressed by power
functional theory and machine-learned force
functionals~\cite{Zimmermann2024NeuralForce}.} For mixtures,
the density of each component can be treated as a species-resolved input
channel, with Cartesian moments formed for each channel and coupled through
CACE products, analogous to the treatment of chemical species in atomistic
CACE~\cite{cheng2024cartesian}. 
Ionic and polar fluids require an additional long-ranged branch: local
molecular field theory already provides one successful separation of learned
short-range correlations from analytical
electrostatics~\cite{Bui2025Ionic}. Alternatively, reciprocal-space,
structure-factor-like features with Coulomb kernels could augment the local
functional, in the spirit of latent Ewald summation~\cite{Cheng2025Latent}. Finally,
the reference density fields need not come from empirical pair potentials.
They can be generated by machine learning interatomic potentials trained on
electronic-structure energies and forces~\cite{keith2021combining}, creating a route from
quantum-mechanical accuracy to mesoscale liquid thermodynamics, as recently
demonstrated for planar neural cDFT~\cite{BuiCox2026AbInitio}.

More broadly, this work brings learned thermodynamic functionals into
three-dimensional continuum-fluid settings. A common scalar
description connects microscopic density organization to response, phase
behavior, solvation and adsorption across distinct geometries.
This combination of three-dimensional equivariance, {grid-resolution transferability}, variational
consistency and supervision from canonical MD generated equilibrium densities
provides a practical foundation for data-driven classical density functionals
of increasingly realistic fluids.

\section{Methods}

\subsection{Equi-cDFT model and implementation}

Each data record is a complete canonical equilibrium one-body density field
$\{\rho_g\}$ on a regular grid, together with the temperature, total particle
number and external potential $V_\mathrm{ext}$ evaluated at the same grid points. The intrinsic
Equi-cDFT model itself takes only $\{\rho_g\}$ and $T$ as inputs; the $V_\mathrm{ext}$ enters the local-balance objective but is not an input to the learned
intrinsic functional. Although the functional is
assembled from local environments, complete fields are processed as individual
examples so that the total free energy and the per-field chemical-potential
offset are well defined.

The ideal-gas integral and all other spatial integrals are evaluated by
voxel-centered quadrature on the density grid, and
canonical minimizations enforce $\Delta V\sum_g\rho_g=N$. Automatic
differentiation is taken with respect to the grid density $\rho_g$, rather than
the voxel occupation $\rho_g\Delta V$; the factors of $\Delta V$ in
Eqs.~\eqref{eq:c1} and \eqref{eq:learned_second_direct_correlation} convert the
result to the corresponding continuum-normalized functional derivatives.

The {Equi-cDFT model} represents the excess free energy as the
sum of a pointwise baseline and a finite-range lattice-CACE correction,
\begin{equation}
\begin{aligned}
F_{\mathrm{exc},\theta}[\rho,T]
&=\Delta V\sum_g\rho_g\Bigl[
a^{\mathrm{loc}}_\theta(\widetilde\rho_g,\widetilde T)\\
&\qquad+a^{\mathrm{CACE}}_\theta
(\widetilde\rho_g,\mathbf B_g,\widetilde T)\Bigr].
\end{aligned}
\label{eq:methods_lattice_cace_energy}
\end{equation}
Here $\widetilde\rho_g$ and $\widetilde T$ denote network inputs normalized by
their corresponding dataset means. The physical, unscaled density $\rho_g$ is
retained in the quadrature.
{The implementation parameterizes $\beta F_{\mathrm{exc},\theta}$.}

For every target voxel, the {representation used here} uses a spherical
neighbor stencil with a cutoff of {four grid spacings at the training
resolution ($2\sigma$)}, containing the {257}
integer offsets satisfying $|\mathbf q|^2\leq {4^2}$. The center is excluded
from the moment sums and supplied separately to the readout, leaving
$N_{\mathrm{nb}}={256}$ noncentral neighbors. We use a single {normalized Gaussian} radial
channel and the 20 raw Cartesian monomials
$q_x^{\ell_x}q_y^{\ell_y}q_z^{\ell_z}$ with total degree
$|\boldsymbol\ell|\leq3$.
{The radial weights are}
\begin{equation}
{w_\alpha(\mathbf q)
=\frac{\exp[-\alpha|\mathbf q|^2]}
{\sum_{0<|\mathbf q'|\leq q_{\mathrm{cut}}}
\exp[-\alpha|\mathbf q'|^2]}.}
\label{eq:methods_radial_weight}
\end{equation}
{The positive exponent was optimized jointly with the network in
logarithmic form.}

Products of the Cartesian moments are retained through correlation order
$\nu=2$ and averaged over the 48 signed axis permutations of the cubic grid,
following the CACE construction~\cite{cheng2024cartesian}. Products
related by cubic symmetry share one
orbit representative, and symmetry-forbidden products vanish. This gives 15
invariant CACE features: two at first order and thirteen at second order. The
CACE readout therefore receives 17 inputs: the center density, the 15
invariants and temperature. Its multilayer perceptron has dimensions
$17\!\rightarrow\!32\!\rightarrow\!16\!\rightarrow\!1$. The pointwise
baseline receives the local density and temperature and uses a
$2\!\rightarrow\!32\!\rightarrow\!16\!\rightarrow\!1$ network. Both
networks use SiLU hidden activations and linear scalar outputs, for a total of
{1,763} trainable parameters.

The model, automatic functional derivatives and equilibrium solvers are
implemented in Python using PyTorch~\cite{Paszke2019PyTorch}. PyTorch automatic
differentiation is applied to the assembled scalar free energy to obtain
$c^{(1)}$ and, when required, $c^{(2)}$. The implementation is publicly
available at \url{https://github.com/BingqingCheng/equicdft}.

{For an $L=8\sigma$ density field, a double-precision evaluation of the
excess free energy and its first functional derivative takes approximately
6.8~ms on one AMD CPU core and 2.6~ms on an NVIDIA L40S GPU at
$h=0.5\sigma$. At $h=0.25\sigma$, the corresponding times are 62~ms and
2.7~ms, respectively.}

\subsection{Training-set generation}
For this study, we use Lennard--Jones reduced units with
$\epsilon=\sigma=m=k_{\mathrm B}=\Lambda=1$. Training data were generated for
a one-component fluid interacting through the
truncated-and-shifted Lennard--Jones (LJTS) potential
\begin{equation}
u(r)=
\begin{cases}
4\epsilon[(\sigma/r)^{12}-(\sigma/r)^6]-u_{\mathrm{LJ}}(2.5\sigma),
&r<2.5\sigma,\\
0,&r\geq2.5\sigma.
\end{cases}
\label{eq:methods_ljts_potential}
\end{equation}
Throughout, $\tau=\sigma\sqrt{m/\epsilon}$ and all simulations use the time
step $0.005\tau$. All canonical MD simulations use a Nos\'e--Hoover
thermostat with damping time $0.5\tau$.

All training simulations used the canonical ensemble in cubic periodic cells
of side $L=8\sigma$. Every trajectory comprised
$4\times10^5$ MD equilibration steps and $4\times10^6$ production steps.
Number density was accumulated every ten production steps, giving 400,000
samples for each equilibrium field.

The external potentials were independently randomized superpositions of
periodic one-, two- and three-dimensional Gaussian components. Their
positions or Cartesian directions, widths, amplitudes and attractive or
repulsive signs were randomized, and a single field could combine components
of different dimensionality. Gaussian widths were typically
$0.2$--$2.0\sigma$, with absolute amplitudes of order {$0.5$--$10\epsilon$} in the energy scale. 

\begin{figure*}[t]
\centering
\includegraphics[width=0.85\textwidth]{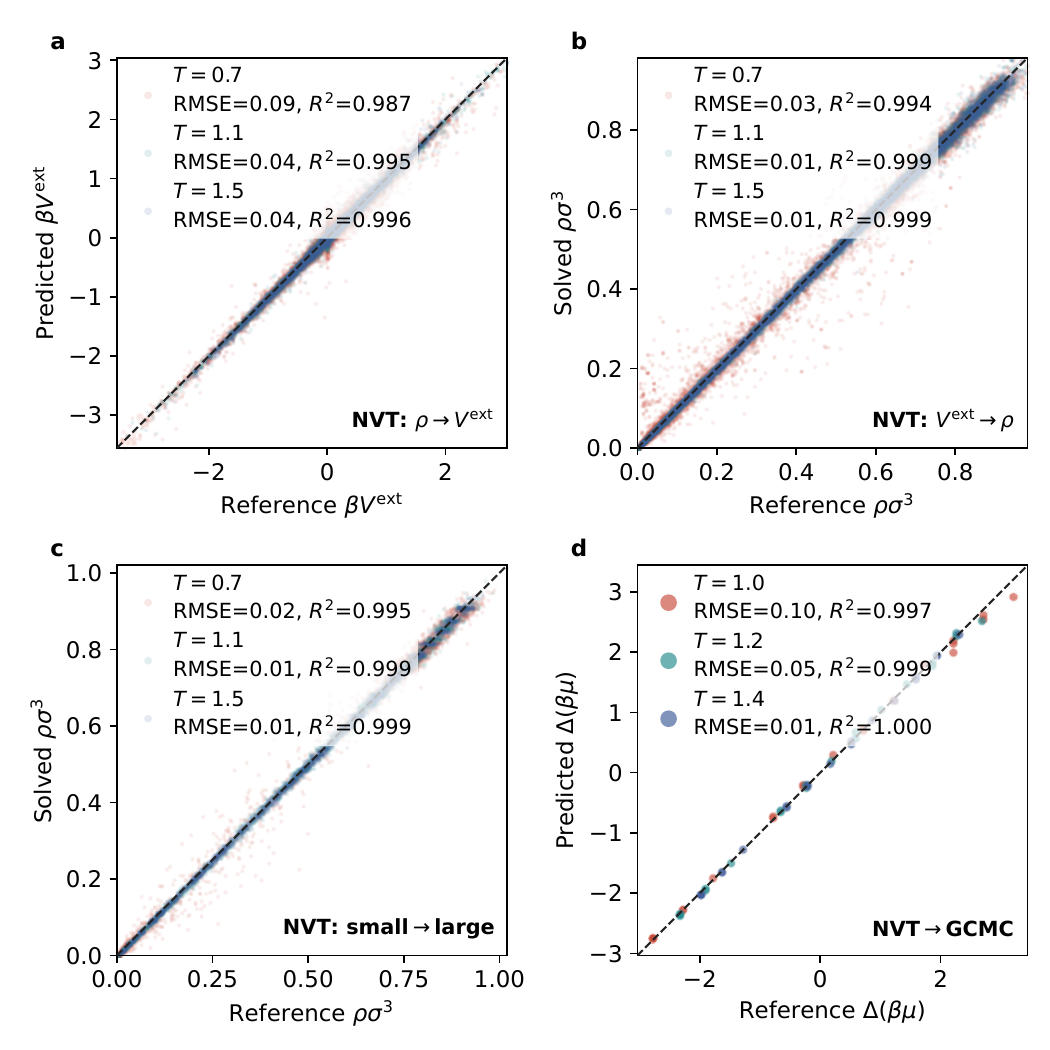}
\caption{\textbf{Combined benchmark of the learned excess free-energy
functional.} 
\textbf{a} Gauge-aligned external fields inferred from {1,590}
held-out NVT equilibrium density fields; one spatially constant gauge is
aligned for each canonical field. 
\textbf{b} Fixed-particle-number {density profiles}
obtained from the external fields for {1,590} held-out fields.
\textbf{c} Transfer of the density solver to {89} test fields in boxes with
3.375 times the training volume. 
\textbf{d} Gauge-aligned dimensionless
chemical-potential differences inferred for all 66 GCMC equilibrium fields by
the NVT-trained functional. Reference and predicted chemical potentials are
centered independently at each temperature, corresponding to one additive
gauge per temperature and no statewise fitting.
{Each dot in panels a--c represents one grid-point reference--prediction
pair. Grid points are subsampled only for display; the reported metrics use
all eligible grid points from the fields.}}
\label{fig:combined-benchmarks}
\end{figure*}

The density fields used for learning were represented on a $16^3$ grid with
spacing $h_0=0.5\sigma$. The analytic external potential was evaluated at the same
grid centers and stored with the time-averaged density. Particle numbers ranged
from $N=8$ to 464, corresponding to whole-box mean densities from
$0.02\,\sigma^{-3}$ to $0.91\,\sigma^{-3}$. {The corpus used for training,
validation and the small-system tests in \figref{fig:combined-benchmarks}
contains 14,066} complete fields and spans the reduced temperatures
$T=0.625$, 0.65, 0.675, 0.7, 0.725, 0.75, 0.775, 0.8, 0.85, 0.9, 0.95,
1.0, 1.05, 1.1, 1.15, 1.2, 1.25, 1.3, 1.4, 1.5, 1.6, 1.7 and 1.8.
{All retained configurations are fluid states.}

All 1,592 fields at $T=0.7$, 1.1 and 1.5 were assigned to the test set to
measure interpolation across temperature.
Fields at these three temperatures are absent from the training and validation set.
Within every remaining source file,
10\% of the fields were assigned to validation. This gives {11,219} training
fields and {1,255} validation fields. Splitting was performed at the
complete-field level.

\subsection{Training}
Equi-cDFT is trained on complete equilibrium density fields using
Eq.~\eqref{eq:chemical_potential_balance_loss}. For each field, the sum and
spatial mean are on grids with
$\rho_g>10^{-3}\sigma^{-3}$. 
We train in single precision with complete-field batches of two. Optimization
uses Adam with an initial learning rate of $10^{-4}$. A
reduce-on-plateau scheduler halves the learning rate after three epochs without
validation improvement, down to a minimum of $10^{-6}$. Training is run for
200 epochs, with checkpoints written every five epochs; the checkpoint with
the lowest validation local-chemical-potential loss is used throughout. On a
single NVIDIA L40S GPU, the {200-epoch fit} required approximately
{2~h 40~min}, or about {48}~s per epoch.

\subsection{\texorpdfstring{{Comparison with a non-equivariant network}}{Comparison with a non-equivariant network}}

\begin{figure}[t]
\centering
\includegraphics[width=\columnwidth]{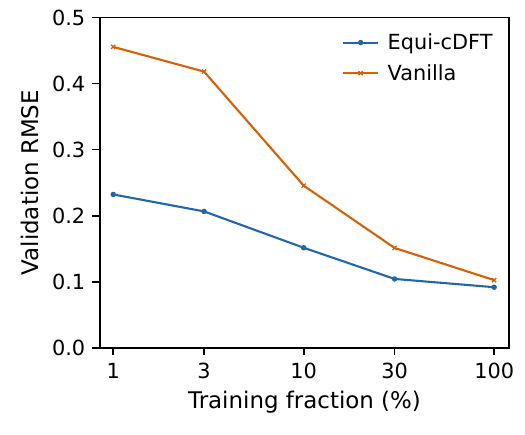}
\caption{{\textbf{Comparison with a non-equivariant voxel network.}
Dimensionless validation RMSE versus training-data fraction for Equi-cDFT
and the voxel network at $h_0=0.5\sigma$. }}
\label{fig:training-data-efficiency}
\end{figure}

{We compared Equi-cDFT with a non-equivariant voxel network that learns
a scalar excess free-energy functional from ordered voxel densities in a
cubic neighborhood of half-width $2\sigma$, together with temperature.
Both architectures were trained using the same chemical-potential-balance
objective on identical nested subsets containing 1\%, 3\%, 10\%, 30\% and
100\% of the training fields, with a fixed validation set.
\figref{fig:training-data-efficiency} shows the dimensionless validation
RMSE of the local chemical-potential imbalance in one training run.}

{Equi-cDFT achieves lower validation error at every training fraction
despite having substantially fewer parameters, demonstrating more efficient
learning for the tested architectures. Additional bulk and interfacial
benchmarks reveal substantially poorer predictions from the non-equivariant voxel network
(Supplementary Fig.~S3).}

\subsection{Benchmark evaluation protocols}

\subsubsection{Held-out forward and inverse tests}
We first test the functional in both directions on NVT holdout fields at
$T=0.7$, 1.1 and 1.5, temperatures omitted entirely from training and
validation. The held-out test set contains {1,590} complete fields in
$L=8\sigma$ cells on $16^3$ grids with spacing $0.5\sigma$, spanning
$N=8$--464.

For the forward test, every held-out density field is supplied to the model.
Automatic differentiation gives $c^{(1)}$ and hence the external-field
prediction
\begin{equation}
\beta V^{\mathrm{ext}}_{\mathrm{raw},g}
=c^{(1)}_{g}-\ln(\rho_g\Lambda^3).
\label{eq:methods_raw_external_field}
\end{equation}
Because canonical training does not determine the additive chemical-potential
gauge, the least-squares constant equal to the active-voxel mean of reference
minus prediction is added independently for each field. 
Here the comparison uses the training mask $\rho_g>10^{-3}\sigma^{-3}$.

The inverse test removes the reference density and supplies only
$V^{\mathrm{ext}}$, $T$ and $N$. 
The adaptive fixed-$N$ Euler solver starts from a uniform density, renormalizes
$N$ exactly after every update and uses adaptive mixing. 

{At $T=0.7$, liquid--vapor domain positions and morphologies can be weakly
constrained, making pointwise density comparisons sensitive to initialization,
which may contribute to the low-temperature scatter.}

Every field contributes uniformly to the displayed parity clouds: 25 active
grid points per field are shown in \figrefsub{a}{fig:combined-benchmarks}, and
25 grid points per field in \figrefsub{b}{fig:combined-benchmarks}. 
The reported metrics use every
eligible grid point from all fields.

\subsubsection{Transfer to larger canonical systems}
To test locality and extensivity across system size, the model trained on
$L=8\sigma$ fields is applied to $L=12\sigma$ boxes
discretized on $24^3$ grids at the same three held-out temperatures. {There
are 29 evaluated fields at $T=0.7$ and 30 each at $T=1.1$ and 1.5, giving
89 fields in total.} They span
$N=32$--1568 and whole-box mean densities
$\rho\sigma^3=0.02$--0.91.

The same fixed-$N$ solver is used as in the inverse test above.
The cloud
in \figrefsub{c}{fig:combined-benchmarks} displays 100 uniformly sampled grid
points from every field, whereas the reported metrics use all {400,896 grid
values at $T=0.7$ and 414,720 at each of $T=1.1$ and 1.5}.

\subsubsection{Grand-canonical chemical-potential differences}
Finally, we assess transfer from canonical training to grand-canonical
chemical-potential differences. The $L=8\sigma$, $16^3$ GCMC test set
contains 22 complete fields at each of $T=1.0$, 1.2 and 1.4, giving 66 fields
in total. The imposed reservoir
chemical potentials span $\mu/\epsilon=-4$ to 2.

For field $a$, the raw prediction is the masked spatial mean
\begin{equation}
\beta\mu^{\mathrm{raw}}_a
=\Bigl\langle
\ln(\rho_g\Lambda^3)+\beta V^{\mathrm{ext}}_g
-c^{(1)}_{g}
\Bigr\rangle_{\rho_g>10^{-3}\sigma^{-3}}.
\label{eq:methods_raw_gcmc_chemical_potential}
\end{equation}
Reference and predicted values are then centered independently at each
temperature,
\begin{equation}
\Delta(\beta\mu)_{a}
=\beta\mu_a-\left\langle\beta\mu\right\rangle_T,
\label{eq:methods_centered_chemical_potential}
\end{equation}
which is one optimal additive $\beta\mu$ gauge per temperature and involves no
statewise fitting. All 22 fields per temperature are evaluated and displayed in
\figrefsub{d}{fig:combined-benchmarks}.

\subsection{\texorpdfstring{{Transfer across grid resolutions}}{Transfer across grid resolutions}}

{We tested whether the functional learned at $h_0=0.5\sigma$ could be
evaluated at different spatial resolutions without retraining. The same
physical fields in periodic $L=6\sigma$ cells were evaluated at
$h=0.125$, 0.25, 0.5 and $1\sigma$. 
The set comprises 69 independently generated mixed-Gaussian
fields at $T=0.8$, 1.0, 1.2 and 1.5 with a maximum grid density $0.95\sigma^{-3}$.
Each field averaged
eight independent NVT replicas, with $4\times10^6$ production steps per replica.
At each spacing, we tested both external-field inference from the
reference density and equilibrium-density recovery by fixed-particle-number
minimization from a uniform density. }

{During transfer, the physical cutoff remains $2\sigma$ and the Gaussian
width is preserved by using $\alpha_h=\alpha_{h_0}(h/h_0)^2$.
Cartesian monomials use the rescaled offsets $(h/h_0)\mathbf q$, retaining
their training-grid units. Network weights and the density and temperature
normalization scales are unchanged.
Details of the descriptor normalization and construction of the paired
reference grids are provided in the Supplementary Information.
\figref{fig:grid-resolution-transfer} compares the same fields at
every resolution. Within the selected
domain, both derivative predictions and equilibrium-density recovery remain
accurate across the tested resolutions, with greater deviations on the
coarsest grid.}

\begin{figure*}[t]
\centering
\includegraphics[width=\textwidth]{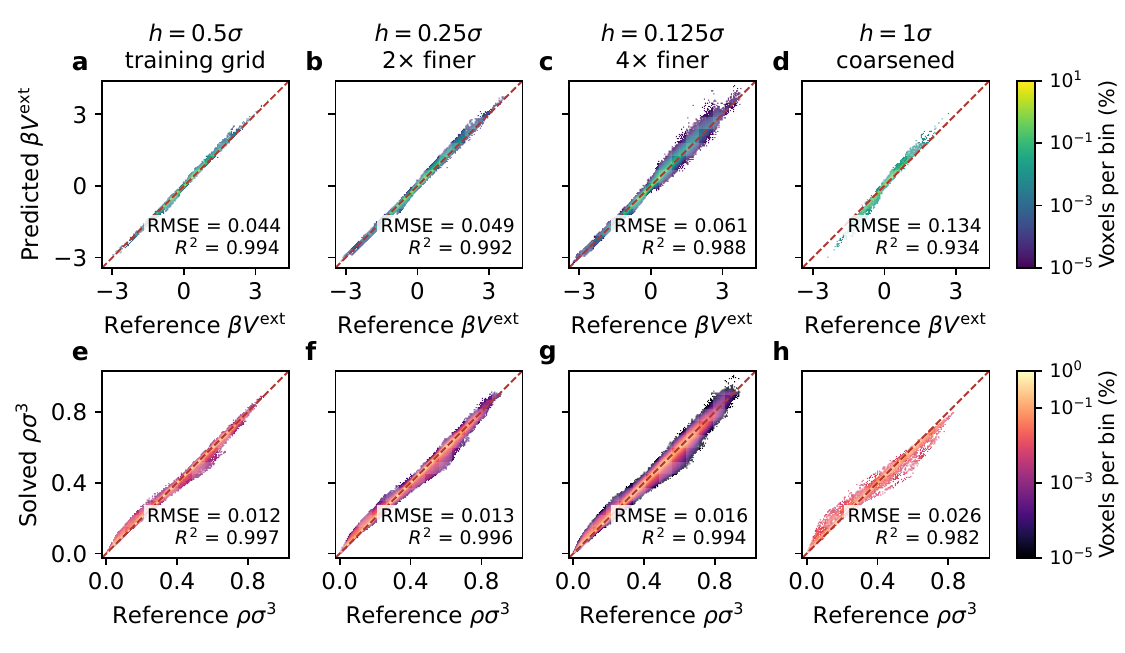}
\caption{{\textbf{Transfer across grid resolutions.}
The Equi-cDFT was trained at $h_0=0.5\sigma$ and evaluated at the four indicated
spacings without retraining. \textbf{a--d} Reference external fields and fields inferred from the
reference density, with one additive canonical constant aligned per field.
\textbf{e--h} Reference densities and predictions from fixed-$N$ minimization.}}
\label{fig:grid-resolution-transfer}
\end{figure*}

\subsection{Interface simulations}

The interfacial references in \figrefsub{d}{fig:bulk-response-phase} were
generated in fully periodic $6\sigma\times6\sigma\times24\sigma$ cells with
$N=276$. Each trajectory used $10^6$ target-temperature equilibration steps
and $4\times10^6$ production steps. Four independent replicas were run at each
temperature, and configurations were stored every 1,000 production steps.
Because a slab can translate along the elongated direction, every production
frame was recentered before averaging. 
The recentered number density was binned with spacing $0.5\sigma$ and averaged
over the transverse plane, all production frames and the four replicas. 

\subsection{Colloid-bridging calculations}
Reference densities and solvent-mediated mean forces were obtained
from canonical MD at two conditions: $T=1.2$, $N=4600$ and $T=0.9$, $N=600$
mobile LJ solvent particles. Both used a periodic cubic box of side
$L=20\sigma$ and the same colloid fields. The solvent particles interacted with one another
through the LJTS pair potential in
Eq.~\eqref{eq:methods_ljts_potential}. The two colloids were not dynamical
particles; each was represented by a fixed analytic one-body field acting on
the LJ solvent. Their centers were
$\mathbf R_L=(L/2-D/2,L/2,L/2)$ and
$\mathbf R_R=(L/2+D/2,L/2,L/2)$. 
Each colloid center generated the following potential on the LJ
solvent particles, using periodic minimum-image distances:
\begin{equation}
\begin{aligned}
\phi(r)&=\frac{H}{1+\exp[(r-R_c)/w_c]}\\
&\quad-\epsilon_a\exp\!\left[-\frac{(r-R_s)^2}{2w_s^2}\right],
\end{aligned}
\label{eq:methods_colloid_field}
\end{equation}
with $R_c=2.6\sigma$, {$w_c=0.25\sigma$}, $H=8\epsilon$,
$R_s=3.4\sigma$, $w_s=0.8\sigma$ and $\epsilon_a=1.4\epsilon$. The total field
was
\begin{equation}
\begin{aligned}
V_{\mathrm{ext}}(\mathbf r;D)
&=\phi(|\mathbf r-\mathbf R_L(D)|)\\
&\quad+\phi(|\mathbf r-\mathbf R_R(D)|).
\end{aligned}
\label{eq:methods_colloid_external_field}
\end{equation}
There was no direct colloid--colloid interaction: $U_{\mathrm{CC}}(D)=0$, so
the reported force is entirely mediated by the LJ solvent.

The $T=1.2$, $N=4600$ scan used one independent trajectory at each
of ten separations, $D/\sigma=4.5,5.0,5.5,6.0,6.5,7.0,7.5,8.0,9.0$ and
10.0. The $T=0.9$, $N=600$ scan used three independent trajectories at each
of twelve separations from $D/\sigma=4.5$ to 10.0 in increments of 0.5.
Every trajectory used 400,000 equilibration steps ($2000\tau$) and 4,000,000
production steps ($20,000\tau$), with the common timestep and thermostat
specified above.

During production, the instantaneous generalized force was sampled every ten
steps. If $f_{x,j}^{(L)}$ and $f_{x,j}^{(R)}$ are the forces exerted on solvent
particle $j$ by the left and right fields, respectively, the measured quantity
was
\begin{equation}
f_D=\frac12\left[
\sum_jf_{x,j}^{(L)}-\sum_jf_{x,j}^{(R)}\right].
\label{eq:methods_colloid_direct_force}
\end{equation}
Each trajectory provided 400,000 force samples separated by
$0.05\tau$, and statistical uncertainties were estimated based on block averaging.
MD free-energy differences were obtained by trapezoidal integration of
$-\langle f\rangle_D$ from $10\sigma$ to each separation, with force errors
propagated through the quadrature weights. 

MD densities were accumulated every ten steps as number densities on
a regular {$80^3$ grid with spacing $0.25\sigma$, with coarse $40^3$ fields obtained by aligned $2\times2\times2$ voxel averaging}. The analytic external field
and $\partial_DV_{\mathrm{ext}}$ were evaluated at the same cell centers, so
the gridded Feynman--Hellmann force was computed without interpolation using
Eq.~\eqref{eq:colloid_feynman_hellmann_grid}, with
{$\Delta V=h^3$ on each grid}.
Equi-cDFT densities were obtained by fixed-$N$ minimization {on $40^3$ ($h=0.5\sigma$)
and $80^3$ ($h=0.25\sigma$) grids.}

\subsection{Porous adsorption calculations}
The gyroid cell containing Lennard--Jones fluid has
$L=16\sigma$, $q=2\pi/L$,
\begin{equation}
g(\mathbf r)=\sin(qx)\cos(qy)+\sin(qy)\cos(qz)
+\sin(qz)\cos(qx),\nonumber
\end{equation}
and the potential on the LJ particles is
\begin{equation}
V_{\mathrm{gyr}}(\mathbf r)=
\frac{8\epsilon}{1+\exp[g(\mathbf r)/{0.20}]}.
\label{eq:methods_gyroid_field}
\end{equation}
The accessible side is $g>0$, and the nominal wall $g=0$ corresponds to
$V_{\mathrm{gyr}}=4\epsilon$.

The structural reference in \figrefsub{a}{fig:gyroid-adsorption}--
\figrefsub{c}{fig:gyroid-adsorption} is a canonical MD simulation at $T=1.2$
with $N=1200$, giving $\langle\rho\rangle\sigma^3=0.29$. The
trajectory used $4\times10^5$ equilibration steps and $4\times10^6$ production
steps. Number density was accumulated every {20} steps on a {$64^3$} grid with
spacing {$0.25\sigma$}, yielding {200,000} sampled fields in the production average.
The matched Equi-cDFT {calculations} used the same field, temperature, box
and total particle number{, with the Equi-cDFT model evaluated on
$h=0.5\sigma$ and $h=0.25\sigma$ grids from uniform initial
densities}. 

The atomistic {reference} values in \figrefsub{d}{fig:gyroid-adsorption} and
\figrefsub{e}{fig:gyroid-adsorption} comprise gyroid and homogeneous
GCMC simulations. 
Each run uses $4\times10^5$ equilibration and $4\times10^6$
production steps using NVE dynamics with a Langevin thermostat
of damping time $0.5\tau$.
Twenty insertion/deletion exchanges were attempted every 50 steps.
{In the pore runs, particle} number and the full {$64^3$, $h=0.25\sigma$} density field were sampled
every 20 steps.
The homogeneous-reservoir GCMC simulations used $L=8\sigma$ and
$V_{\mathrm{ext}}=0$ at the same chemical potentials. 

Equi-cDFT was solved at fixed
$N$ from a uniform initial density, using the same temperature, grid, box and
external field. The chemical potential was
obtained by averaging the local chemical potential. The bulk branch was evaluated on uniform fields. For comparison with the GCMC chemical-potential reference, one
constant offset, determined from the mean reference-minus-prediction
difference across the bulk states, was applied unchanged to both geometries.
The adsorption curve was constructed by interpolating the two
chemical-potential--density branches and matching them at equal chemical
potential; the common offset cancels from this matching.

\section*{Data availability}
{The trained model, numerical source data, evaluation and application scripts,
and complete training, validation, test and reference datasets are available in
the \texttt{lj-paper-v2} release at
\url{https://github.com/BingqingCheng/equicdft-lj-data/releases/tag/lj-paper-v2}.}

\section*{Code availability}
The Equi-cDFT library is publicly available at
\url{https://github.com/BingqingCheng/equicdft}.

\section*{Acknowledgements}
Research reported in this publication was supported by the National Institute
of General Medical Sciences of the National Institutes of Health under Award
Number R35GM159986.

OpenAI Codex assisted with literature organization, manuscript editing, and the development and review of plotting and figure-generation scripts. The author directed its use and independently verified all scientific claims, calculations, code, figures, and references.

\section*{Author contributions}
BC designed and performed the study and wrote the paper.

\section*{Competing interests}
The author declares no competing interests.

\end{document}